\documentstyle[12pt]{article}

\begin{document}
\centerline {\bf {\LARGE Spin polarised nuclear matter and its}}
\smallskip
\centerline {\bf {\LARGE application to neutron stars}}
\bigskip
\centerline {\bf V.S. Uma Maheswari, D.N. Basu, J.N. De }
\smallskip
\centerline {\bf {\it Variable Energy Cyclotron Centre,}}
\centerline {\bf {\it 1/AF, Bidhan Nagar, Calcutta - 700064, India.}}
\medskip
\centerline { and }
\centerline {\bf S.K. Samaddar }
\smallskip
\centerline {\bf {\it Saha Institute of Nuclear Physics,}}
\centerline {\bf {\it 1/AF, Bidhan Nagar, Calcutta - 700064, India.}}
\vskip 1.0 true cm

\noindent {\bf Abstract}\\
An equation of state(EOS) of nuclear matter with
explicit inclusion
of a spin-isospin dependent force
is constructed from a finite range, momentum
and density dependent effective interaction. This EOS is found to be in
good agreement with those obtained from more sophisticated models
for unpolarised nuclear matter.
Introducing spin degrees of freedom,
it is found that at density about 2.5 times
the density of normal nuclear matter the neutron matter undergoes
a ferromagnetic transition.
The maximum mass and the radius of the neutron star agree favourably
with the observations.
Since finding quark matter rather than
spin polarised nuclear matter at the core of neutron stars
is more probable,
the proposed EOS is also applied to the study of hybrid stars.
It is found using the bag model picture that
one can in principle describe both the mass and size as well as the
surface magnetic field of hybrid stars satisfactorily.
\vfill
\noindent {\bf PACS Number(s):\ 97.60.Jd, 21.65.+f}

\newpage

\noindent {\bf 1.\ Introduction }
\vskip 0.5 true cm

Since the pioneering
works of Baade and Zwicky\cite{bz} and Oppenheimer and Volkoff\cite{ov}
about half a century ago,
and the identification of pulsars as rotating neutron stars around the  year
1968\cite{gold}, several studies have been made
to unravel the mystery of the structure of
the neutron stars.
It is a well accepted fact that the density inside a neutron
star varies from the surface to the core by about 15 orders of magnitude.
Understanding the structure of such complex objects requires
an accurate knowledge
of the equation of state(EOS) of neutron star matter
in the different density
regions. The extremely low density domain and the subnuclear region can be
well described by the EOS given
by Feynmann-Metropolis-Teller (FMT)\cite{fmt}
and
Baym-Pethick-Sutherland (BPS)\cite{bps} respectively.
The EOS of the dense nuclear
matter is still riddled with some uncertainties. A consistent EOS should
describe the compressional properties\cite{jpb}
of nuclei near the ground state and
also of the hot and dense nuclear material
that is created in energetic nuclear
collisions\cite{sg}. It should also answer such
important questions as whether a star
at the later stage of its life explodes or
not\cite{geb} and how the neutron star is
born. Renewed interest in the
rapid cooling of neutron stars\cite{cjp,jml} by the direct
URCA process has also an immediate bearing on the nuclear EOS.

In the non-relativistic framework,
EOS of nuclear matter have been constructed
with phenomenological effective interactions or
from realistic interactions with
different degrees of sophistication[11-16].
Most of these EOS can explain the
properties of neutron stars like their mass,
size, moment of inertia etc. well
within the observational limits.
However, in regard to the magnetic properties
of neutron stars, till now no acceptable
explanation exists for the origin of
the rather large magnetic field $(\sim 10^{12} G)$ at the surface.
 A recent
analysis\cite{mag} of
binary millisecond pulsars suggests that a permanent
component of this magnetic
field could exist, sustained by a spontaneous magnetised phase inside the
neutron
star. Attempts have been made to explain the
presence of the magnetic field by
means of a ferromagnetic transition.

In the framework of Hartree-Fock theory, employing hard and also soft core
potentials, Pfarr\cite{pf} does not get such a ferromagnetic transition.
A similar conclusion is reached by
Forseth and Ostgaard\cite{forseth} who made the
calculations in the lowest order constrained
variational method of Pandharipande
using soft core potentials; with a hard core potential a transition to the
ferromagnetic state was seen to
occur though at $\sim 30 \ \rho_o$,
where $\rho_o$ is the normal nuclear matter
density.
In a relativistic $\sigma+ \omega$ Hartree-Fock approach, a ferromagnetic
transition is also predicted by Niembro et al\cite{niembro} at too high a
density. However, in an improved model\cite{marcos} with inclusion of
$\pi$ and $\rho$ mesons in addition
to $\sigma + \omega$, a ferromagnetic transition
is seen to occur at a comparatively much lower density, $\rho \sim 3.5\rho_o$,
but the incompressibility of normal nuclear matter is found to be too high
($\sim 450$ MeV).
It would therefore be interesting to know whether a ferromagnetic
phase transition is possible at a density realisable in neutron star matter
with an EOS
with firmer grounds in the
experimental realities of normal nuclear matter and finite nuclei.
This is the primary motivation of this work.

There is a strong possibility that at the core densities of neutron
stars, there is a phase transition from nuclear matter to quark
matter[22-25].
About a decade ago,
Witten conjectured\cite{witten} that the strange quark matter(SQM) might be
the absolute ground state of hadronic matter
i.e., the mass energy per baryon may be less than 930 MeV.
If this is true, then the possibility that the pulsars
are rotating strange quark stars may not be ruled out.
Even if SQM is not the absolute ground state (i.e. at densities less than
the hadron-quark transition density the
nuclear matter is energetically favoured than SQM),
one may still find hybrid stars having quark cores with nucleon
envelops. Since our understanding of the confinement/deconfinement of quarks
is
far from complete, all the aforesaid possibilities are only speculative.
One should also keep in mind
that the theoretical framework used in general to study the phase
transition is phenomenological and simplistic in nature.

In view of all the above, we would like to investigate in this paper
the following.
Firstly, whether there exists an EOS which consistently describes the
nuclear matter
and finite nuclear properties,
as well as predict a ferromagnetic transition at a
density realisable in the interior of neutron stars.
Secondly, whether the same
EOS which predicts a ferromagnetic transition permits a hadron-quark
phase transition
and thereby the formation of a hybrid star.
And finally using such an EOS, whether we can
consistently explain the structural properties such as mass, radius and moment
of inertia as well as the presence of the magnetic field at the surface within
the same model.
\\

\noindent {\bf 2.\ Theoretical framework }
\vskip 0.5 true cm

In the following, we briefly outline the procedure to obtain the nuclear
equation of state in a non-relativistic framework and discuss its
merits and limitations.

\noindent {\bf {\it 2.1.\ Equation of state}}

The phenomenological momentum and density dependent finite range interaction
employed here to obtain the equation of state is a modified version of
Seyler-Blanchard interaction\cite{db}.
To treat spin-polarised isospin asymmetric
nuclear matter, the interaction has been generalised to include explicitly
the spin-isospin dependent channel.
The interaction between two nucleons with separation $ r $
and relative momentum $ p $ is given by,
\begin{equation}
v_{eff} (r,p,\rho ) = -C_{\tau s} \left [ 1-{p^2\over b^2}-
d^2{\left ( \rho_1(r_1)+\rho_2(r_2) \right )}^n \right ]{e^{-r/a}\over {r/a}}
\ ,
\end{equation}
where $ a $ is the range and $ b $ defines the strength of the repulsion in the
momentum dependence of the interaction.
The parameters $ d $ and $ n $ are measures of the strength of the
density dependence,
and $\rho_1$ and $\rho_2$ are the densities at the sites
of the two interacting nucleons. The subscripts $\tau $ and $ s $ in the
strength parameter $C_{\tau s }$
refer to the likeness $ l $ and the unlikeness $ u $ in the isotopic spin
and spin of the two nucleons respectively; for example, $C_{ll}$ refers to
interactions between two neutrons or protons with parallel spins, $C_{lu}$
refers to that between neutrons or protons with opposite spin {\it etc}.
The energy per nucleon $E$ and the pressure $P$
in the mean-field approximation can then be worked out\cite{db} as
\begin{equation}
E=
{1\over \rho}
\sum_{\tau s} \rho_{\tau s} \left [ T {J_{3/2} (\eta_{\tau s})
\over J_{1/2} (\eta_{\tau s} ) } (1-m^*_{\tau s} V_{\tau s}^1 ) +
{1\over 2} V_{\tau s}^0 \right ]
\ ,
\end{equation}
\begin{equation}
P=\sum_{\tau s} \rho_{\tau s}
\left [ {2\over 3}T {J_{3/2} (\eta_{\tau s})
\over J_{1/2} (\eta_{\tau s} ) } + V_{\tau s}^0 +
{1\over 2}b^2
\left (1-d^2(2\rho )^n \right ) V_{\tau s}^1 + V_{\tau s}^2
\right ] \ .
\end{equation}
Here,
$J_k (\eta )$ are the Fermi integrals, $V_{\tau s}^0$ and $V_{\tau s}^2$
are the single-particle and the
rearrangement
potentials, $V_{\tau s}^1$ is the coefficient of the quadratic
momentum dependent term in the potential
and defines
the effective mass $m^*_{\tau s}$, $T$ the temperature, and $\eta$
is the fugacity given by
$\eta_{\tau s} = (\mu_{\tau s} - V_{\tau s}^0 - V_{\tau s }^2 )/T $.
For the unpolarised nuclear matter(NM),
the expressions for $V_{\tau s}^0$ {\it etc} are given in
 ref.\cite{db}. It is straightforward to extend these to the case of
polarised nuclear matter and are as given below:
\begin{eqnarray}
V_{\tau s}^0  &=& -4\pi a^3 \left ( 1-d^2 {(2\rho_o)}^n \right )
\left [ C_{ll}\rho_{\tau ,s}+C_{lu}\rho_{\tau,-s}+C_{ul}\rho_{-\tau,s}+
C_{uu}\rho_{-\tau,-s} \right ] \nonumber \\
\quad & & + {8\pi^2 a^3 \over b^2 h^3} \
{\Large \{ }
C_{ll} {(2m_{\tau,s}^*T)}^{5/2} J_{3/2}(\eta_{\tau,s})
+ C_{lu} {(2m_{\tau,-s}^*T)}^{5/2} J_{3/2}(\eta_{\tau,-s})
\nonumber \\
&\quad &  + C_{ul} {(2m_{-\tau,s}^*T)}^{5/2} J_{3/2}(\eta_{-\tau,s})
      + C_{uu} {(2m_{-\tau,-s}^*T)}^{5/2} J_{3/2}(\eta_{-\tau,-s})
{\Large \} },
\nonumber \\
V_{\tau s}^1 &=& {4\pi a^3\over b^2}
\left [ C_{ll}\rho_{\tau ,s}+C_{lu}\rho_{\tau,-s}+C_{ul}\rho_{-\tau,s}+
C_{uu}\rho_{-\tau,-s} \right ], \nonumber \\
V_{\tau s}^2
&=& 4\pi a^3 d^2 n{(2\rho_o)}^{n-1}\sum_{\tau^{\prime},s^{\prime}}
\left [ C_{ll}\rho_{\tau^{\prime} ,s^{\prime}}
      + C_{lu}\rho_{\tau^{\prime} ,-s^{\prime}}
      + C_{ul}\rho_{-\tau^{\prime} ,s^{\prime}}
      + C_{uu}\rho_{-\tau^{\prime} ,-s^{\prime}}
\right ]\ \rho_{\tau^{\prime},s^{\prime}}, \nonumber \\
m_{\tau s}^*
&=& { \left [ {1\over m_{\tau }} + 2V_{\tau s}^1 \right ] }^{-1}.
\end{eqnarray}

One usually defines the neutron and proton spin excess parameters ( spin
asymmetry ) as
\begin{eqnarray}
\alpha_{n} &=&
({\rho_{n \uparrow}-\rho_{n \downarrow}})/\rho \ , \nonumber \\
\alpha_{p} &=&  ({\rho_{p \uparrow}-\rho_{p \downarrow}})/\rho \ ,
\end{eqnarray}
where
\begin{equation}
\rho = \rho_n +\rho_p = (\rho_{n\uparrow}+\rho_{n\downarrow}) +
			(\rho_{p\uparrow}+\rho_{p\downarrow})\ ,
\end{equation}
is the number density. We then define the proton fraction as
$x=\rho_p/\rho$. It is related to the isospin asymmetry parameter $X$
as,
\begin{equation}
X=(1-2x) = (\rho_n - \rho_p )/\rho \ .
\end{equation}
We also define
the spin excess parameter as $Y=\alpha_n+\alpha_p$ and the
spin-isospin excess parameter as $Z=\alpha_n - \alpha_p $.
One can then express the energy per nucleon $E/A$ of the NM
at zero temperature as
\begin{equation}
E/A =E_{V}+E_{X} X^2 +E_{Y} Y^2 +E_{Z} Z^2
\ ,
\end{equation}
where terms higher
than those quadratic in $X$, $Y$ and $Z$ are neglected. Here
$E_V$ is the volume
energy of the symmetric nuclear matter, taken as $-16.1$ MeV
and $E_X$ is the usual
symmetry (isospin) energy, taken to be $34.0$ MeV. The
quantities $E_Y$ and $E_Z$ are the spin and the spin-isospin symmetry
energies of the NM respectively. Their values are uncertain to some extent.
We take\cite{ph,arh}
$E_Y = 31.5$ MeV and $E_Z=35.0$ MeV in conformity with the generalised
hydrodynamical model of Uberall\cite{ub},
where ${( E_Z/E_X )}^{1/2} \simeq 1.1 $.
In terms of the
strength parameters $C_{\tau s}$, the volume and the symmetry
energies are written in the form,
\begin{equation}
{ \left (
\begin{array}{cccc}
-A&-A&-A&-A\\
-B&-B&C&C\\
-B&C&-B&C\\
-B&C&C&-B
\end{array}
\right ) }
{ \left (
\begin{array}{c}
C_{ll}\\C_{lu}\\C_{ul}\\C_{uu}
\end{array}
\right ) }
=
{ \left (
\begin{array}{c}
E_V-{3p_F^2/(10m)}\\
E_X-{p_F^2/(6m)}\\
E_Y-{p_F^2/(6m)}\\
E_Z-{p_F^2/(6m)}
\end{array}
\right ) }
\ .
\end{equation}
Here $p_F$ is the Fermi momentum of the one-component nuclear matter
corresponding to the density $\rho$ of the polarised NM,
$A=\alpha (\beta -\delta )$, $B=\alpha ( \beta -20\delta /9)$,
$C=\alpha ( \beta - 10\delta/9)$, $\alpha = {8\pi^2a^3p_F^3/(3h^3)}$,
$\beta = 1-d^2{(2\rho)}^n$ and $\delta={6p_F^2/(5b^2)}$.
For a fixed value of $n$,
the parameters $C_{\tau s}$, $a$, $b$ and $d$ are then determined by
reproducing $E_V$, $E_X$, $E_Y$, $E_Z$, the saturation density of
normal nuclear matter$(\rho_0 = 0.1533\ fm^{-3}) $, the surface energy
coefficient ($a_S = 18.0$ MeV ), and the energy dependence of the real
part of the nucleon-nucleus optical potential.
The parameter $n$ is determined by reproducing the breathing-mode
energies\cite{mmm}.

The parameters of the interaction are listed below:
\begin{eqnarray}
C_{ll} = -305.2\ {\rm MeV} &\quad& a=0.625\ {\rm fm} \nonumber \\
C_{lu} = \ 902.2\ {\rm MeV} &\quad& b=927.5\ {\rm MeV/c} \nonumber \\
C_{ul} = \ 979.4\ {\rm MeV}  &\quad& d=0.879\ {\rm fm^{3n/2}} \nonumber \\
C_{uu} = \ 776.2\ {\rm MeV}  &\quad & n=1/6 .\nonumber
\end{eqnarray}
With the above value of the parameter $n$, the
incompressibility of symmetric nuclear matter is $K=240 $ MeV.
\\

\noindent {\bf 2.2.\ {\it Merits and limitations}}
\vskip 0.5 true cm

It has been tested that the above interaction reproduces
quite well
the ground state binding energies, root mean square charge radii,
charge distributions and giant monopole resonance energies for a host of
even-even nuclei ranging from $^{16}O$ to very heavy systems.
Interactions of this type have been used before with great success
by Myers and Swiatecki\cite{myers} in the context of nuclear
mass formula.
We have also seen that
for symmetric nuclear matter, our results agree extremely
well with those calculated in a variational approach by
Friedmann and Pandharipande(FP)\cite{fp} with $v_{14}+TNI $ interaction in
the density range ${1\over 2}\rho_0 \le \rho \le 2\rho_0 $.
However, for unpolarised pure neutron matter, the energies calculated
with our interaction are somewhat higher compared to the FP energies,
particularly at higher densities. The entropy per particle for
neutron matter calculated with our interaction at different
temperatures agrees extremely well with the corresponding
FP results. In Fig.1, the energy per particle for neutron matter
is displayed as a function of density at zero temperature.
For comparision, the FP energies\cite{fp} and those obtained with the
Bethe-Johnson(BJ) potential\cite{adj} in a sophisticated correlated
basis function approach are also displayed.
The BJ curve is very close to ours for neutron matter.
This good agreement between our calculations and those
reported in
refs.\cite{fp} and \cite{adj} suggests that the present interaction
can be extrapolated with some confidence to neutron matter or to
nuclear matter with large isospin asymmetry at high densities.
It can also be mentioned that such an interaction satisfies
the Landau-Migdal stability criteria\cite{rudra}.

All the well-known non-relativistic nuclear equations of state suffer from
lack of causality at high densities. The velocity of sound in nuclear
matter then becomes superluminal. The effective interaction used by us is
no exception.
In Fig.2, we have plotted the velocity of sound in units of $c$
as a function of the ratio $\rho/\rho_o$ in the case of neutron matter
taking $\alpha_n=0,\ 0.3,\ {\rm and}\ 0.5$.
It can be seen that as spin-polarisation
increases, the EOS becomes softer and the velocity of sound $v_s$ becomes
acausal only at increasingly higher densities. This superluminous behaviour
of $v_s$, particularly for the unpolarised neutron matter,
suggests that the extrapolation of such an EOS to very high nuclear
densities may not be advisable.
\\

\noindent {\bf 3.\ Ferromagnetic phase transition }
\vskip 0.5 true cm

It would be interesting to investigate whether the nuclear EOS discussed
in the previous section, in addition to the consistent description of the
nuclear matter and finite nuclear properties predicts a ferromagnetic
transition at densities meaningful in the context of neutron stars.

It has been conjectured\cite{baym,pines} that in the neutron star matter,
in contrast to nuclei where the neutrons generally pair up to spin $J=0$
in their ground states, the neutrons may pair up to spin $J=1$ at higher
nuclear densities,
thus leading to a ferromagnetic transition. To investigate
this aspect, we calculate the energy per particle for the spin polarised
neutron matter with a representative value of $\alpha_n = 0.5$ and compared
with that calculated for unpolarised neutron matter in Fig.1.
We find that above $\rho \sim 2.5\rho_o$, the energy of polarised matter
is lower compared to that for unpolarised neutron matter. This reflects
that the neutrons that pair up to $J=0$ at lower densities undergo a
transition to a spin polarised configuration as density builds up.
In otherwords,
the system prefers a ferromagnetic state for $\rho > 2.5\rho_o$.

The behaviour of the magnetic susceptibility $\chi$ of neutron matter as
a function of density also portrays the occurence of ferromagnetic
phase transition. In general, the magnetic susceptibility is
defined\cite{huang} as $\chi = {\partial {\it M}\over \partial H}$, where
$H$ is the magnetic field and
${\it M}= \mu_n (N_{\uparrow}-N_{\downarrow})/V $
is the total magnetisation
per unit volume with $\mu_n$ being the magnetic moment of a neutron.
Using the definition of $\alpha_n$[Eq.(5)], we rewrite ${\it M}$ as
${\it M} = \mu_n \alpha_n \rho_n $.

We need to determine the optimum value of $\alpha_n$ using the
energy minimisation criteria,
\begin{equation}
{\partial (E_H (\rho, \alpha_n)/N )\over \partial \alpha_n }
\mid_{\alpha_n = \alpha_n^0} = 0.
\end{equation}
The total energy $E_H/N$ per particle of a system of $N$ number of neutrons
in the presence of an external weak magnetic field $H$ is,
\begin{equation}
E_H (\rho, \alpha_n )/N = E (\rho, \alpha_n )/N - \left ( \mu_n H \alpha_n
\right ) .
\end{equation}
Expanding the energy $E(\rho, \alpha_n )$ in powers of $\alpha_n$ upto
$O(\alpha_n^2)$, we get
\begin{eqnarray}
E(\rho, \alpha_n )/N &=& E(\rho, \alpha_n=0)/N + {1\over 2}\alpha_n^2
{\partial^2 (E (\rho, \alpha_n )/N)\over \partial \alpha_n^2 }
\mid_{\alpha_n=0}, \nonumber \\
\quad &\equiv & e_0 + {1\over 2}\alpha_n^2 e_2.
\end{eqnarray}
Because the energy $E(\rho, \alpha_n)$ is symmetric in $\alpha_n$, all
the odd derivatives in the expansion of $E(\rho, \alpha_n )$ vanish.
(It may be said here that only in the calculation of $\chi$, we have
expanded $E(\rho, \alpha_n)$ in powers of $\alpha_n$, otherwise we
have calculated it numerically.)
Then, the optimum value $\alpha_n^0$ is determined by minimising
the energy[Eq.(11)] as $\alpha_n^0= \mu_n H /e_2 $.
Now, we can determine $\chi$ and it is given as,
\begin{equation}
\chi = {\partial \over \partial H} \left ( \mu_n \alpha_n^0 \rho_n \right )
= {\mu_n^2 \rho_n \over e_2 }.
\end{equation}
Using the effective interaction given in Eq.(1), we get
\begin{eqnarray}
e_2 &=& {\partial^2 (E (\rho, \alpha_n )/N)\over \partial \alpha_n^2 }
\mid_{\alpha_n=0}, \nonumber \\
 & = & -2\pi a^3 \rho_n
\left [ A_1(C_{ll}-C_{lu}) - {20\ 2^{2/3}\over 9}\ A_2 (2C_{ll}-C_{lu})
\right ] + {2^{2/3} p_F^2\over 3m},
\end{eqnarray}
where $A_1= 1-d^2{(2\rho_n)}^n $ and $A_2= \rho_n p_F^3/b^3 $.
In the limit of no interaction, $e_2$ is simply given by the
kinetic term alone, i.e. $e_2^{free} = {2^{2/3} p_F^2/(3m)}$.
It is then straightforward to calculate the ratio $\chi_{free}/\chi$,
where $\chi_{free}$ is the magnetic susceptibility of the non-interacting
neutron gas. The onset of a ferromagnetic transition is depicted by the
vanishing of the ratio $\chi_{free}/\chi$.
Further, the effect of the nuclear matter incompressibility
$K$ on the ferromagnetic transition
density is studied and the results are shown in Fig.3.
As the value of $K$ is increased from 240 MeV to 304 MeV
(by increasing the density exponent $n$ of the effective
interaction given by eq.(1) from $1/6$ to $2/3$),
the density at which the transition takes place
decreases from $\sim 2.4 \rho_o$ to $\sim 2.3 \rho_o $.
It is thus seen that the effective
interaction given in Eq.(1) predicts a ferromagnetic phase transition
at a density $\rho \sim 2.4\rho_o$.

In previous calculations in the non-relativistic formalism using
realistic interactions, one usually does not find such a ferromagnetic
transition \cite{pf,forseth} or if there is such a possibility, it occurs at
densities\cite{forseth} not realisable in the context of neutron stars.
In a relativistic framework, one finds that the simple $\sigma \ + \ \omega$
 model\cite{niembro} does not suggest any such transition. However, with an
 improved
model\cite{marcos} including other mesons like
$\rho$ and $\pi$ in addition to
$\sigma+\omega$, one finds a transition at about $\rho \sim 3.5\rho_o$.
But, the incompressibility of the normal nuclear matter obtained
in this model is too high $(\sim 450\ MeV)$.
In contrast, our nuclear interaction that predicts a ferromagnetic transition
at a similar density yields a value of incompressibility that is close
to the one well-accepted\cite{jpb}.
\\

\noindent {\bf 4.\ Structural properties of neutron stars}
\vskip 0.5 true cm

In this section, we explore the various static properties of neutron stars
such as proton fraction, mass, size and moment of inertia using the
proposed equation of state. We further study the influence of spin
polarisation on these observables.
\\

\noindent {\bf 4.1.\ {\it Beta-equilibrium proton fraction }}
\vskip 0.5 true cm

In recent years, attention has been drawn to the direct URCA process in
neutron stars which may be the primary mechanism for its rapid cooling.
This can, however, occur only when the beta-equilibrium proton fraction $x$
in the star is $\geq 0.11$,
where only electrons are considered, and $\geq$ 0.148, if both electrons and
muons are considered.
It would be interesting to know whether spin polarisation favours
or disfavours direct URCA process.
In our study,
the lepton energy per particle $E_L(\rho,x)$ is given by the relativistic,
ideal Fermi-gas expression\cite{shapiro};
in addition to $e^-$, $\mu^-$ are also considered as and
when they are energetically favoured.
At beta-equilibrium, one has
${\partial \over \partial x} \left ( E(\rho,x)+E_L(x) \right )=0$,
where $E(\rho, x)$ is the baryonic
energy  per particle including the rest masses.
In Fig.4, the beta-equilibrium proton fraction
thus obtained in the neutron star matter is displayed as a function of the
baryon density invoking the condition of charge neutrality.
The upper panel  corresponds
to unpolarised matter, and the lower panel displays  that for spin
polarised matter with $\alpha_n=0.3$ and $\alpha_n=0.5$.
From the upper panel, it can be seen that for $\alpha_n =0$, $x$ shows a peaked
structure against density.
When only $e^-$ are considered, $x$ increases as density increases, reaches
a maximum value of about 0.085 at $\rho \simeq 3\rho_o$, and then decreases
to very low values at higher densities.
With the inclusion of $\mu^-$,
the structure of the curve remains almost the same; however, it lies
higher than that of the former case, at all densities. The peak value is then
about 0.11. The lower panel of Fig.4 displays proton fractions for the spin
polarised neutron star matter with $\alpha_n=0.3$ and $\alpha_n =0.5$.
Here both muons and electrons
are taken into consideration for calculating proton fractions at
beta-equilibrium. With increasing spin polarisation, the proton fraction
becomes smaller at any density. It may be noted that the present EOS
in use does not favour direct URCA process, since the proton fraction is
always below the critical value. Introduction of spin polarisation disfavours
direct URCA process even more. It may be remarked that various calculations
give different conclusions\cite{jml} regarding the direct URCA process.
It may also be mentioned that inclusion of exotic processes like pion
condensation or kaon condensation in dense neutron star matter may enhance
the proton fraction thereby favouring direct URCA process\cite{mittet},
 but occurence
of such exotic phenomena is still very much unsettled\cite{cjp}.
\\

\noindent {\bf 4.2.\ {\it Mass and size of neutron stars}}
\vskip 0.5 true cm

We now determine the structure of neutron stars using a composite EOS
i.e. FMT, BPS, Baym-Bethe-Pethick(BBP)\cite{bbp} and the present interaction
with progressively increasing densities.
Then, the total mass and the size of the neutron star can be obtained
by solving the general relativistic
Tolman-Oppenheimer-Volkoff(TOV) equation,
\begin{equation}
{dP(r)\over dr} = -{G\over c^4} \
{\left [\epsilon (r)+P(r)\right ] \left [ m(r)c^2 + 4\pi r^3 P(r) \right ]
\over { r^2 \left [ 1-{2Gm(r)\over rc^2} \right ] }},
\end{equation}
where,
\begin{equation}
m(r)c^2 = \int_0^r \epsilon (r^{\prime}) d^3r^{\prime }.
\end{equation}
The quantities $\epsilon (r)$ and $P(r)$ are the energy density and pressure
at a radial distance $r$ from the centre, and are given by the
equation of state. The mass of the star contained within a distance $r$
is given by $m(r)$.
The size of the star is determined by the boundary condition $P(R)=0$
and the total mass $M$ of the  star integrated upto the surface $R$
is given by $M=m(R)$. The single integration constant needed to solve
the TOV equation is $P_c$, the pressure at the center of the star
calculated at a given central density $\rho_c$.

The mass functions of the star thus obtained as a function of its central
density are shown in Fig.5 for four different values of spin polarisation
$(\alpha_n =0.0,\ 0.3, \ 0.4, \ 0.5)$.
The radii, central densities and surface redshifts
corresponding to the maximum mass $M_{max}$
configuration are tabulated in Table 1 for three values of $\alpha_n$.
The surface redshift $z_s$ is defined\cite{bbp} as
\begin{equation}
z_s = {\left [ 1-{2GM\over Rc^2} \right ]}^{-1/2} - 1.
\end{equation}
Values of $M_{max}/M_{\odot}$ are also given in the same table.
It can be seen that the maximum mass and the corresponding radius
and surface redshift $z_s$ decrease
with increasing polarisation. On the other hand, the central density
pertaining to $M_{max}$ configuration increases as $\alpha_n$ increases.
This is because of the fact that the spin polarised neutron star matter is
more compressible than the unpolarised one.
The measured mass\cite{rj} of $4U 0900-40$,
$(1.85\pm 0.3)M_{\odot}$ possibly
provides the lower limit of the maximum mass of the neutron star. Our
calculations with the values of $\alpha_n$ taken are well within this limit.
The maximum mass of neutron star obtained from calculations with $\alpha_n=0.6$
and above are found to be below the present observational limits and
hence we have restricted our calculations upto $\alpha_n=0.5$.
It may be mentioned that the unpolarised neutron matter is close to being
superluminal at the central density corresponding to the maximum mass.
For polarised neutron star matter though the central density increases
significantly, the sound velocity is always luminal because of the
softness of the polarised matter towards compression.

We have also studied the sensitivity of the mass distribution $ m(r) $
to $\alpha_n$, where $ m(r) $ denotes the total mass contained within a given
radial distance $ r $ from the centre. Fig.6 displays
$m(r)/M_{max}$ as a function
of the density $\rho (r)$ at that point $ r $, where $M_{max}$
corresponds to the maximum mass
configuration at a given $\alpha_n$.  The zero of the abscissa refers to
the surface
of the neutron star; the points where the mass functions $m(r)$ meet the
abscissa
refer to the centre of the star. It is found that nearly $90\%$ of the
mass of the
neutron star lies at densities higher than $3\rho_o$.
\\

\noindent {\bf 4.3. \ {\it Moment of inertia of neutron stars}}
\vskip 0.5 true cm

The moment of inertia of neutron stars is calculated by assuming the
star to be rotating slowly
with an uniform angular velocity $\Omega$\cite{arnett}.
The angular velocity ${\bar \omega}(r)$ of a point in the star measured
with respect to the angular velocity of the local inertial frame is
determined by the equation,
\begin{equation}
{1\over r^4}{d\over dr}\left [ r^4 j {d{\bar \omega}\over dr} \right ]
+ {4\over r}{dj\over dr} {\bar \omega} =0,
\end{equation}
where,
\begin{equation}
j= e^{-\phi (r)} {\left ( 1-{2Gm(r)\over rc^2} \right ) }^{1/2}.
\end{equation}
The function $\phi (r)$ is constrained by the condition,
\begin{equation}
e^{\phi (r) } \mu (r) = {\rm constant } = \mu (R) {\sqrt { 1-{2GM\over Rc^2}}},
\end{equation}
where the chemical potential $\mu (r) $ is defined as,
\begin{equation}
\mu (r) = { \epsilon (r) + P(r) \over \rho (r) }.
\end{equation}
Using these relations, Eq.(18) can be solved subject to the boundary
conditions that ${\bar \omega }(r)$ is regular as $r \longrightarrow 0$,
and ${\bar \omega} (r) \longrightarrow \Omega $ as $r \longrightarrow \infty $.
Then moment of inertia of the star can be calculated using the definition
$I=J/\Omega $, where the total angular momentum $J$ is given as
\begin{equation}
J = {c^2\over 6G}\ R^4 \ {\left ( {d{\bar \omega}\over dr} \right ) }\mid_{r=R}.\end{equation}
Values of $I$ thus obtained for three values of $\alpha_n$ are plotted as a
function of the central density in Fig.7.
It can be seen that the maximum value of $I$ greatly depends on $\alpha_n$.
As $\alpha_n$ is increased from 0.0 to 0.5, $I_{max}$ has decreased
by about 50$\%$.

	We have thus calculated the important structural properties of
neutron stars and studied their dependence on the spin polarisation factor.
It is found that upto a value $\alpha_n = 0.5$, the properties obtained
here are well within the observational limits. Motivated by this success
in explaining the structural properties of neutron stars, we are curious
to see how far we can account for the surface magnetic field.
Our present study suggests that the neutron star matter at densities
$\rho > 2.5\rho_o$ is spin polarised. Taking $\alpha_n=0.5$, one
immediately realises that the surface magnetic field is largely
overestimated$(\sim 10^{16}G)$.
As noted earlier, it is probable that
that at core densities one finds quark matter rather than spin polarised
nuclear matter. Therefore, we in the following section explore this
plausibility and its implications upon both the structural and magnetic
properties of stars.
\\

\noindent {\bf 5. \ Hybrid stars}
\vskip 0.5 true cm

Here, we construct the $\beta-$ equilibrated, electrically neutral
quark matter equation of state and
employ it to understand the structural properties
of the hybrid stars,i.e. quark cores with nucleon envelopes.

The equation of state of a three flavour quark matter consisting $u$, $d$
and $s$ quarks is obtained here using the phenomenological MIT
bag model\cite{bag}.
The total kinetic energy density of a system of non-interacting, relativistic
quarks of flavour $\tau $ and mass $m_{\tau }$ is given as,
\begin{equation}
\epsilon_{\tau } = {3\over 8\pi^2 }\ {{(m_{\tau }c^2)}^4\over {(\hbar c)}^3}
\left [ x_{\tau }{\sqrt {1+x_{\tau}^2}}\ (1+2x_{\tau }^2) -
ln(x_{\tau} +{\sqrt {1+x_{\tau}^2}}) \right ],
\end{equation}
where $x_{\tau }=p_F^{\tau}/(m_{\tau}c)$, $p_F^{\tau}$ being the Fermi
momentum and is related to the quark number density $\rho_{\tau }$ of
a given flavour as $p_F^{\tau} = \hbar {(\pi^2 \rho_{\tau})}^{1/3}$.
The densities pertaining to the three flavours can be expressed in terms of the
total quark number density $\rho_q$ and the asymmetry parameters
$\delta_{ud}$ and $\delta_{us}$ as:
\begin{eqnarray}
\rho_u &=& (\rho_q/3) \left [ 1-\delta_{ud}-\delta_{us} \right ], \nonumber \\
\rho_d &=& (\rho_q/3) \left [ 1+2\delta_{ud}-\delta_{us} \right ], \nonumber \\
\rho_s &=& (\rho_q/3) \left [ 1-\delta_{ud}+2\delta_{us} \right ],
\end{eqnarray}
where $\delta_{ud} = (\rho_d-\rho_u)/\rho_q$,
$\delta_{us} = (\rho_s-\rho_u)/\rho_q$ and
$\rho_q=\rho_u+\rho_d+\rho_s$.
Similarly, the energy density $\epsilon_L$ pertaining to a system of
relativistic non-interacting electron gas can be calculated\cite{shapiro}.

We then determine the equilibrium composition of
the quark matter subject to the $\beta-$
equilibrium condition,
\begin{equation}
\mu_d - \mu_u = \mu_e \quad {\rm and} \quad \mu_d = \mu_s,
\end{equation}
and the charge neutrality condition,
\begin{equation}
\rho_e = {1\over 3} \left ( 2\rho_u - \rho_d - \rho_s \right ).
\end{equation}
Using Eq.(24) one obtains,
$\rho_e = -(\rho_q/3)\ (\delta_{ud}+\delta_{us})$.
Similarly, we can express the chemical potentials $\mu_u$, $\mu_d$
and $\mu_s$ in terms of the three quantities $\rho_q$,
$\delta_{ud}$ and $\delta_{us}$ as follows:
\begin{eqnarray}
\mu_u &=& \left ( {\partial \epsilon_q \over \partial \rho_u }\right )
{}_{\mid_{\rho_d,\rho_s}} = {\partial \epsilon_q\over \partial \rho_q}
-{1+\delta_{ud}\over \rho_q}\ {\partial \epsilon_q\over \partial \delta_{ud}}
-{1+\delta_{us}\over \rho_q}\ {\partial \epsilon_q\over \partial \delta_{us}},
\nonumber \\
\mu_d &=& \left ( {\partial \epsilon_q \over \partial \rho_d }\right )
{}_{\mid_{\rho_u,\rho_s}} = {\partial \epsilon_q\over \partial \rho_q}
+{1-\delta_{ud}\over \rho_q}\ {\partial \epsilon_q\over \partial \delta_{ud}}
-{\delta_{us}\over \rho_q}\ {\partial \epsilon_q\over \partial \delta_{us}},
\nonumber \\
\mu_s &=& \left ( {\partial \epsilon_q \over \partial \rho_s }\right )
{}_{\mid_{\rho_u,\rho_d}} = {\partial \epsilon_q\over \partial \rho_q}
-{\delta_{ud}\over \rho_q}\ {\partial \epsilon_q\over \partial \delta_{ud}}
+{1-\delta_{us}\over \rho_q}\ {\partial \epsilon_q\over \partial \delta_{us}},
\end{eqnarray}
where $\epsilon_q = \sum_{\tau} \epsilon_{\tau } + B$ is the total
quark energy density and $B$ is the bag parameter. Using these expressions,
the $\beta-$ equilibrium conditions can be rewritten as,
\begin{eqnarray}
{\partial \epsilon_q\over \partial \delta_{ud}}
-{\partial \epsilon_q\over \partial \delta_{us}} &=& 0,
\nonumber \\
{2 \over \rho_q}\ {\partial \epsilon_q\over \partial \delta_{ud}}
+{1 \over \rho_q}\ {\partial \epsilon_q\over \partial \delta_{us}}
&=& \mu_e,
\end{eqnarray}
where $\mu_e = {\sqrt {p_{F,e}^2c^2+m_e^2c^4}}$.
Thus, for a given baryon density $\rho_b = \rho_q/3$, the three quantities
$\rho_e$, $\delta_{ud}$ and $\delta_{us}$ are fixed by the Eqs.(26) and (28).
Subsequently, the equation of state is completely described by the
total energy density $\epsilon_{QM}$ and the pressure $P_{QM}$ of the
system calculated for a given $\rho_b$ using the definitions,
\begin{eqnarray}
\epsilon_{QM} &=& \sum_{\tau} \epsilon_{\tau} (\rho_q, \delta_{ud},\delta_{us})
+\epsilon_{L} (\rho_q, \delta_{ud},\delta_{us}) +B , \nonumber \\
P_{QM} &=& \rho_q \ {\partial \epsilon_{QM}\over \partial \rho_q} -
\epsilon_{QM}.
\end{eqnarray}
It may be mentioned that in our present study we have omitted the lowest
order quark-quark interaction terms as it can be
effectively absorbed into the bag constant\cite{berger}.
The masses of the quarks are taken to be: $m_u=5$ MeV, $m_d=10$ MeV
and $m_s=200$ MeV.

To know whether there is a phase transition from the nuclear matter
to quark matter, we compare the total energies per baryon obtained
in the two phases. The EOS corresponding to NM is given by Eqs.(2-3)
and that of QM is given by Eq.(29).
In Figs.8 and 9 we show the energy per baryon as a function of the
baryon density for two values of bag constant,
$B^{1/4}=155$ MeV and $B^{1/4}=170$ MeV respectively. These are compared with
the curves obtained using $\alpha_n=0.0$ and $\alpha_n=0.5$ in the same figures.
The baryon density $\rho_b$ at which the polarised$(\alpha_n=0.5)$ and an
unpolarised$(\alpha_n=0.0)$ curves intersect is denoted by $\rho_{\small FM}$.
For the nuclear interaction given in Eq.(1), $\rho_{\small FM} \simeq 0.405\
fm^{-3}$.
Similarly, the baryon density at which either of the NM curves intersect
with the QM one is denoted by $\rho_{\small HQ}$.

It can be seen in Fig.8 that for $B^{1/4}=155$ MeV, the unpolarised NM
is energetically favoured upto a density $\rho_b \simeq 0.255\ fm^{-3}$.
As $\rho_b$ is increased further, the quark matter is found to be the
lowest energy state. Therefore, there is no region of spin polarised NM
in this particular case. On the other hand, for $B^{1/4}=170$ MeV, it can
be seen that for densities within the values $\rho_{\small FM} \simeq
0.405\ fm^{-3}$ and $\rho_{\small HQ} \simeq 1.05\ fm^{-3}$, the
polarised NM is the state of lowest energy.
Thus, in this case(Fig.9), one has a spin polarised region sandwitched between
a quark matter core$(\rho_b \geq \rho_{\small HQ})$ and a unpolarised nuclear
matter envelope$(\rho_b \leq \rho_{\small FM})$.
Further, we also studied the dependence of the
width of the spin polarised region, roughly given by
$(\rho_{\small HQ}-\rho_{\small FM})$, on the strange quark mass.
In Fig.10, we have plotted
$(\rho_{\small HQ}-\rho_{\small FM})/\rho_{\small HQ}$
as a function of the bag paramter for three values of $m_s$.
Negative values of $(\rho_{\small HQ}-\rho_{\small FM})/\rho_{\small HQ}$
indicate that there is no region of spin polarised NM inside a hybrid star.
In otherwords, to have a spin polarised region sandwitched between an
unpolarised NM envelope and a QM core, $\rho_{\small HQ}$ must be
greater than $\rho_{\small FM}$. It can be seen that for allowed
values of $B^{1/4}$ and $m_s$, the presence and non-presence of a polarised
region are equally probable.
Thus,
it is clear from the above discussions that the allowed range of $B$ and $m_s$
values is not able to decide upon the presence/non-presence of a spin polarised
region in the hybrid star.
It would be interesting to know whether the observational limits on the
mass and the size of pulsars would help in finding an answer.

In view of this, we explore the structural properties of the
hybrid stars choosing three particular values of the bag parameter $B$ while
keeping the strange quark mass $m_s$ fixed at 200 MeV.
The three different configurations corresponding to the three sets of
MIT bag model parameters are shown schematically in Fig.11. It is clear
from the figure that for a fixed value of $m_s$,
one can appropriately choose the value of $B$ so
that the calculated surface magnetic field is of the order of the
observed value$(\sim 10^{12}\ G)$.
The mass, size, central density and surface redshift were then
calculated for the three values of $B$. The results obtained
pertaining to the maximum mass configuration are given in Table 2.
It can be seen that $M_{max}/M_{\odot}$ and $R$ decreases as the
value of $B$ is increased. For $B^{1/4}=170$ MeV, $M_{max}/M_{\odot}$
is found to be less than the observed value. Therefore, one may
conclude that large regions of spin polarised matter in a hybrid
star is ruled out by the observational limit on the mass of pulsars.
But, our present study cannot decide upon either of the two
configurations obtained with $B^{1/4}=155$ MeV and $B^{1/4}=164$ MeV,
since the mass of the respective stars and their radii are well within
the acceptable limits.
We would however like to mention that using the proposed equation of
state and the MIT bag model with acceptable parameters($B^{1/4} \sim 164$
 MeV, $m_s$=200 MeV), one can in principle describe both the
structural properties as well as the surface magnetic field
satisfactorily.
\\

\noindent {\bf 6.\ Summary }
\vskip 0.5 true cm

To summarise, we have constructed a nuclear equation of state from a
finite range momentum and density dependent interaction and then
applied it to investigate some properties of spin polarised
nuclear matter with particular reference to the neutron star
matter. The parameters of the interaction have a firm basis
in the well known properties of nuclear matter and of finite nuclei.
Extrapolating this interaction to neutron matter and to higher
densities, it is found that the present equation of state agrees
well with those obtained from more sophisticated calculations.

Introducing spin degrees of freedom, it is seen that at density
$\rho \sim 2.5\rho_o$, the neutron star matter undergoes a
ferromagnetic phase transition. This aspect is demonstrated
by studying the density
behaviour of the total energy(Fig.1) and the magnetic susceptibility(Fig.3).
To the best of our knowledge, this is the first
non-relativistic calculation that gives a ferromagnetic
transition at a density well realisable in the
neutron star core.

The proposed equation of state is then applied to the
investigation of the structure of neutron stars.
With increasing polarisation, the maximum mass and
the corresponding radius decreases whereas the
value of central density increases. The
maximum mass of stars and their radii obtained
from calculations with $\alpha_n \leq 0.5$ are found to be
well in agreement with the observations.

There is a good possibility that one finds quark matter rather
than spin polarised nuclear matter at the core of stars.
We therefore investigated this plausibility using the
MIT bag model. We found that using the proposed EOS and
the MIT bag model, one can in principle obtain the
maximum mass, size and the surface magnetic field
of hybrid stars well within the acceptable limits.

\newpage
\centerline {\bf FIGURE CAPTIONS }
\vskip 1.0 true cm
\noindent {\bf Fig.1.}\ The energy per particle for pure neutron matter
as a function of density is plotted for $\alpha_n=0$ and $\alpha_n=0.5$.
The calculated results of Friedman-Pandharipande\cite{fp} and with
Bethe-Johnson potential\cite{adj} are also shown.
\vskip 0.5 true cm
\noindent {\bf Fig.2.}\ Velocity of sound $v_s$ obtained in units of
$c$ for pure neutron matter taking
$\alpha_n=0.0, \ 0.3 \ {\rm and} \ 0.5$ is
plotted as a function of the density ratio $\rho/\rho_o$, where
$\rho_o=0.1533\ {\rm fm^{-3}}$ is the symmetric nuclear matter saturation
density.
\vskip 0.5 true cm
\noindent {\bf Fig.3.}\ Magnetic susceptibility $\chi$ of
pure neutron matter calculated for two values
of nuclear matter incompressibility $K$
is shown as a
function of the density $\rho $, where $\chi_{free}$ is the
magnetic susceptibility of non-interacting neutron gas.
\vskip 0.5 true cm
\noindent {\bf Fig.4. }\ Beta-equilibrium proton fraction $x$ as a
function of density of neutron star matter. In the upper panel, results
are shown for $\alpha_n=0$ with $e^-$ and $e^-+\mu^-$ considered for
$\beta-$ equilibrium. In the lower panel, the results for
$\alpha_n=0.3$ and $\alpha_n=0.5$ are shown with $e^-+\mu^-$.
\vskip 0.5 true cm
\noindent {\bf Fig.5.}\ The neutron star mass is plotted as a function
of central density.
\vskip 0.5 true cm
\noindent {\bf Fig.6.}\ The integrated mass upto radius $r$ is plotted
as a function of the density at radius $r$ for different spin
polarisation $\alpha_n$.
\vskip 0.5 true cm
\noindent {\bf Fig.7} \ Moment of inertia obtained using
$\alpha_n$=0.0, 0.3 and 0.5 is shown as a function of
density $\rho $, where
$\rho_o=0.1533\ {\rm fm^{-3}}$ is the symmetric nuclear matter saturation
density.
\vskip 0.5 true cm
\noindent {\bf Fig.8} The total energy per baryon of quark matter(QM)
calculated using the bag model picture is compared with the energies
per baryon of unpolarised$(\alpha_n=0.0)$ and polarised$(\alpha_n=0.5)$
nuclear matter(NM).
\vskip 0.5 true cm
\noindent {\bf Fig.9} Same as Fig.8, but for $B^{1/4}$=170 MeV.
\vskip 0.5 true cm
\noindent {\bf Fig.10} The width of the spin polarised region,
roughly given by
$(\rho_{\small HQ}-\rho_{\small FM})/\rho_{\small HQ}$, is plotted
as a function of the bag parameter $B$ for three values of
strange quark mass, $m_s$=100, 200 and 300 MeV.
Negative values of
$(\rho_{\small HQ}-\rho_{\small FM})/\rho_{\small HQ}$
imply that there is no spin polarised region inside a star.
\vskip 0.5 true cm
\noindent {\bf Fig.11} Schematic representation of the three
configurations considered in our study of hybrid stars.
\vskip 1.0 true cm
\centerline {\bf TABLE CAPTIONS }
\vskip 1.0 true cm
\noindent {\bf Table 1}\ Values of the mass $M_{max}$, size $R$, central
density $\rho_c$, surface redshift $z_s$ and moment of inertia $I$
obtained in the case of neutron stars using three values of
$\alpha_n$ are shown. The tabulated values correspond to the maximum
mass configuration.
\vskip 1.0 true cm
\noindent {\bf Table 2}\ Values of the mass $M_{max}$, size $R$, central
density $\rho_c$, surface redshift $z_s$ and moment of inertia $I$
obtained in the case of hybrid stars using three values of bag
parameter $B$ are shown. The strange quark mass $m_s$=200 MeV.
The tabulated values correspond to the maximum
mass configuration.
\vfill
\newpage
\centerline {\bf Table 1}
\vspace {0.5in}
\begin{center}
\begin{tabular}{|c|c|c|c|c|}
\hline
\multicolumn{1}{|c|}{$\alpha_n$} &
\multicolumn{1}{|c|}{$M_{max}/M_{\odot}$} &
\multicolumn{1}{|c|}{$R$} &
\multicolumn{1}{|c|}{$\rho_c/\rho_o$} &
\multicolumn{1}{|c|}{$z_s$} \\
\multicolumn{1}{|c|}{} &
\multicolumn{1}{|c|}{} &
\multicolumn{1}{|c|}{$(km)$} &
\multicolumn{1}{|c|}{} &
\multicolumn{1}{|c|}{} \\
\hline
$0.0$ & 2.03 & 10.3 & 7.6 & 0.55\\
$0.3$ & 1.89 & 9.9 & 8.4 & 0.52\\
$0.5$ & 1.53 & 8.7 & 11.7 & 0.44\\
\hline
\end{tabular}
\end{center}
\vskip 0.5 true cm
\centerline {\bf Table 2}
\vspace {0.5in}
\begin{center}
\begin{tabular}{|c|c|c|c|c|}
\hline
\multicolumn{1}{|c|}{$B^{1/4}$} &
\multicolumn{1}{|c|}{$M_{max}/M_{\odot}$} &
\multicolumn{1}{|c|}{$R$} &
\multicolumn{1}{|c|}{$\rho_c/\rho_o$} &
\multicolumn{1}{|c|}{$z_s$} \\
\multicolumn{1}{|c|}{(MeV)} &
\multicolumn{1}{|c|}{} &
\multicolumn{1}{|c|}{$(km)$} &
\multicolumn{1}{|c|}{} &
\multicolumn{1}{|c|}{} \\
\hline
$155$ & 1.62 & 10.1 & 8.8 & 0.38\\
$164$ & 1.46 & 9.3 & 10.3 & 0.37\\
$170$ & 1.37 & 8.8 & 11.4 & 0.36\\
\hline
\end{tabular}
\end{center}
\end{document}